\documentstyle[prl,aps,multicol,epsfig]{revtex}

\begin{document}

\title{Critical Point of a Weakly Interacting Two-Dimensional Bose Gas}

\author{ Nikolay Prokof'ev$^{1}$, Oliver Ruebenacker$^{1}$, and
Boris Svistunov $^{2}$}

\address{
$^1$ Department of Physics, University of Massachusetts,
Amherst, MA 01003, USA \\
$^2$ Russian Research Center ``Kurchatov Institute", 123182 Moscow,
Russia}

\maketitle
\begin{abstract}
We study the Berezinskii--Kosterlitz--Thouless transition in a
weakly interacting 2D quantum Bose gas using the concept of
universality and numerical simulations of the classical
$|\psi|^4$-model on a lattice. The critical density and chemical
potential are given by  relations $n_c=(mT/2\pi \hbar^2) \ln(\xi
\hbar^2/ mU)$ and  $\mu_c=(mTU/\pi \hbar^2) \ln(\xi_{\mu} \hbar^2/
mU)$, where $T$ is the temperature, $m$ is the mass, and $U$ is
the effective interaction. The dimensionless constant $\xi= 380
\pm 3$ is very large and thus any quantitative analysis of the
experimental data crucially depends on its value. For $\xi_{\mu}$
our result is $\xi_{\mu} = 13.2 \pm 0.4 $. We also report the
study of the quasi-condensate correlations at the critical point.

\end{abstract}


\begin{multicols}{2}

\narrowtext An accurate microscopic expression for the critical
temperature of the Berezinskii--Kosterlitz--Thouless (BKT)
transition \cite{BKT} has been a weak point of the theory of
weakly interacting two-dimensional Bose gas. The theory of Ref.~
\onlinecite{Popov} (see also \cite{KSS,Fisher} and analysis
below), suggests that the critical density of the BKT transition
in the weakly interacting system reads (we set $\hbar=1$)
\begin{equation}
n_c = {mT \over 2 \pi} \ln {\xi \over mU}  \; .
\label{n_c_quantum}
\end{equation}
However, the value of $\xi$ cannot be obtained within standard
analytical treatments since $\xi$ is related to the system
behavior in the fluctuation region where the perturbative
expansion in powers of $U$ does not work. With unknown $\xi$, one
finds Eq.~(\ref{n_c_quantum}) rather inaccurate unless $mU$ is
exponentially small. Moreover, as we will find in this Letter, the
value of $\xi$ is very large: $\xi \approx 380$. This
means that for all experimentally available up to date (quasi-)2D
weakly interacting Bose gases \cite{Safonov,Ketterle}
the  quantitative analysis of the data for the
critical ratio $n_c/T_c$ requires a precise value of $\xi$.
In the system of spin-polarized atomic hydrogen on helium film
\cite{Safonov}, the value of $mU$ is of order unity \cite{KSS};
in the recently created quasi-2D
system of sodium atoms \cite{Ketterle}, $mU$ is of order
$10^{-2}$, according to the formula of Ref.~\onlinecite{Petrov}.

To quantitatively describe the BKT transition in a weakly
interacting Bose gas, it is sufficient to solve a classical-field
$|\psi|^4$-model with the effective long-wave Hamiltonian
\cite{Popov}
\begin{equation}
H[\psi]= \int \left\{  {1 \over 2m}|\nabla \psi|^2 + {U \over 2}
|\psi|^4 - \mu' |\psi|^2 \right\} \, d {\bf r} \; , \label{H}
\end{equation}
where $\mu' $ is the chemical potential, and $\psi$ is the classical
complex field.

In this Letter, we first discuss the origin of the relation
(\ref{n_c_quantum}) in the limit of small $U$, and how quantum and
classical models relate to each other. Then we present our numeric
results (for the critical density, critical chemical potential,
and quasi-condensate correlations at the BKT point) obtained by
simulating the critical behavior of the 2D $|\psi|^4$-model on a
lattice using recently developed Worm algorithm \cite{Worm} for
classical statistical models. In particular, we show that the
quasi-condensate correlations are very strong at $T_c$, in
agreement with the experimental observation in the spin-polarized
atomic hydrogen \cite{Safonov} and quantum Monte Carlo simulations
\cite{KKKPS}.

A simple dimensional analysis of the Hamiltonian (\ref{H}) allows
to write a generic formula for the critical point in a weakly
interacting 2D $|\psi|^4$-model. The routine itself is completely
analogous to that in the 3D case (see, e.g., \cite{Baym,KPS}), but
final results naturally reflect the specifics of the 2D case.

We begin with introducing the mode-coupling momentum, $k_c$, that
characterizes the onset of strong non-linear coupling between the
long-wave harmonics of $\psi ({\bf r})$ (harmonics with $k \gg
k_c$ are almost free). This momentum is just the inverse of the
{\it healing} length, or vortex core radius, $r_c$\, \cite{BKT}.
We denote by $\tilde{n}$ the contribution to the total density due
to strongly coupled harmonics, and introduce the renormalized
chemical potential
\begin{equation}
\tilde{\mu} = \mu ' - 2 U \int_{k>k_c} n_k^{\rm (ideal)} d^2 k
/(2\pi)^2   \label{mu_k0}
\end{equation}
by subtracting the mean field contribution of non-interacting
high-momentum harmonics. Here
$n_k=\langle \, |\psi_{\bf k}|^2 \rangle$, and
$\langle \ldots \rangle$ stands for the statistical averaging.

An estimate for $\tilde{n}$ follows from the Nelson-Kosterlitz
formula
\begin{equation}
n_s = {2 m T \over \pi} \; , \label{NK}
\end{equation}
since it is intuitively expected that $ \tilde{n} \sim n_s$. An
independent estimate of the parameters of the fluctuation region
is obtained by considering when all three terms in Eq.~(\ref{H})
are of the same order:
\begin{equation}
k_c^2/m \sim |\tilde{\mu}| \sim \tilde{n} U \; , \label{rel1}
\end{equation}
and relating $ \tilde{n} \sim \sum_{k<k_c} n_k \sim k_c^2 n_{k_c}$
to the renormalized chemical potential by using
$T/|\tilde{\mu}|$ in place of the occupation number $n_{k_c}$.
By definition, $k_c$ separates strongly coupled and free harmonics, and thus
$n_{k_c} \sim T/[ k_c^2/2m - \tilde{\mu}]  \sim T/ |\tilde{\mu}|$.
The final order-of-magnitude estimates read (at $T=T_c$)
\begin{eqnarray}
\tilde{n}      &\sim  & m T \;,           \label{n_tilde} \\
 k_c           &\sim  & m (UT)^{1/2} \; , \label{k_c} \\
 \tilde{\mu}   &\sim  & U m T \;.         \label{mu_tilde}
\end{eqnarray}

We are now in a position to derive Eq.~(\ref{n_c_quantum})
for the critical density. In 2D the main contribution
to the integral
\begin{equation}
n =  \int  n_k  d^2 k/(2 \pi)^2 \; , \label{rel4}
\end{equation}
comes from large momenta between $k_c$ and
some ultra-violet scale $k_*$. The value and physical meaning of $k_*$
depend on the model. For classical lattice
models $k_*$ is given by the inverse lattice spacing;
in the continuous quantum system $k_* \sim \sqrt{mT}$
is the thermal momentum. At $k_c \ll k \ll k_*$
we have $n_k \approx 2mT/k^2$, and thus can write
\begin{equation}
n_c = {mT \over 2\pi} \ln (C k_*^2 /k_c^2 ) \; ,
\label{n_c}
\end{equation}
where $C$ is some constant. Critical density,
Eq.~(\ref{n_c_quantum}), for the  quantum Bose gas is obtained by
substituting $C k_*^2/k_c^2 \equiv \xi mT/m^2UT = \xi/mU $.

The dependence on the ultraviolet cutoff is associated with the
properties of {\it ideal} systems only, while the long-wave
behavior of all weakly-interacting $|\psi |^4$-theories is
universal. This fact allows one to relate results for different
models by adding and subtracting non-interacting contributions,
i.e., up to higher order corrections in $U$ the difference between
models ${\cal A}$ and ${\cal B}$ is given by $(n_c^{\cal (A)} -
n_c^{\cal (B)}) = \int [n_k^{\rm (ideal~{\cal A})} - n_k^{\rm
(ideal~{\cal B})}] \, d^2k/(2\pi)^2$. In what follows, the
reference system ${\cal A}$ will be the classical lattice model
with lattice spacing $a$, and our results are analyzed using
\begin{equation}
n_c^{\rm (lat)} = {m T \over 2 \pi} \ln {A \over m^2a^2UT}  \; .
\label{n_c_lattice}
\end{equation}
The actual system of interest is the quantum Bose gas, so we add
and subtract the corresponding ideal-system contributions to get
\begin{equation}
   \ln {A \over \xi ma^2T} = {1 \over 2\pi mT } \left(
\int_{BZ} { T ~d^2 k\over E({\bf k}) }  -
\int {d^2 k  \over e^{k^2/2mT} -1 } \right)
\; , \label{A_xi}
\end{equation}
where $BZ$ means that the first integral is over the Brillouin
zone, and $E({\bf k})$ is the dispersion law for the ideal lattice
model such that $ E({\bf k} \to 0) \to  k^2/2m$. [The divergences
of the two integrals in Eq.~(\ref{A_xi}) at $k \to 0$ compensate
each other.]

Our simulations were done for the simple square lattice
Hamiltonian
\begin{equation}
H = \sum_{{\bf k} \in BZ} [E({\bf k}) - \mu ] |\psi_k |^2 +
 {U\over 2 } \sum_i |\psi_i|^4\; ,
\label{H_lat}
\end{equation}
where $\psi_k$ is the Fourier transform of the
complex lattice field  $\psi_i$, and
\begin{equation}
E({\bf k}) = (1 / ma^2) [ 2-  \cos (k_xa) -  \cos (k_ya) ]
\label{E}
\end{equation}
is the tight-binding dispersion law.
With this dispersion relation the r.h.s.
in (\ref{A_xi}) can be evaluated analytically and we obtain
the ``conversion'' formula
\begin{equation}
\xi = A / 16 \; .
\label{xi_A}
\end{equation}
Since final results for dimensionless constants do not depend on
$m$, $T$, and $a$, in numerical simulations
we set $a=1$, $T=1$, and $m=1/2$ for convenience.

The above consideration for the critical density can be readily
generalized to the critical chemical potential, with the result

\begin{equation}
\mu_c ={m T U \over \pi} \ln {\xi_{\mu}  \over mU } \; .
\label{mu_c}
\end{equation}
First, we notice that Eq.~(\ref{mu_c}) immediately follows form
Eqs.~(\ref{mu_tilde}) and  Eq.~(\ref{mu_k0}) because the
mean-field term is proportional to $ - (mUT/\pi ) \ln (mU)$ (we
actually deal with exactly the same integral). Since the
renormalized value $\tilde{\mu }$ is universal, to account for the
difference between the classical and quantum models one has to add
and subtract mean-field contributions dominated by the ideal
behavior. Thus, if the classical model is analyzed using $\mu_c=(m
T U / \pi ) \ln [ A_{\mu} / m^2a^2 UT ] $, one has to apply
$\xi_{\mu} = A_{\mu} / 16$ to get the quantum result,
Eq.~(\ref{mu_c}).

We now turn to our numerical procedure. To simulate the
grand-canonical Gibbs distribution corresponding to the
Hamiltonian (\ref{H_lat}), we employ the Worm algorithm (see
Ref.~\cite{Worm} for the description) that has demonstrated its
efficiency for the analogous problem in 3D \cite{KPS}. The formal
criterion of the critical point for the finite-size system is
based on the exact (Nelson--Kosterlitz) relation (\ref{NK}): We
say that the system of linear size $L$ is at the critical point,
if its superfluid density, $n_s(L)$, satisfies $n_s(L) = 2 m T /
\pi$. [The superfluid density has a direct estimator in the Worm
algorithm via the statistics of winding numbers \cite{Worm}, and
its autocorrelation time does not suffer from critical slowing
down.]

The finite size scaling of $n_c(L)$ is well known from the
Kosterlitz-Thouless renormalization group theory \cite{BKT}
\begin{equation}
n_c(L) = n_c - \frac{A'\, m T}
{\ln^2  \big[ A''\, L m (UT)^{1/2} \big] } \; ,
\label{rel5}
\end{equation}
where $A'$ and $A''$ are dimensionless constants.
A similar relation applies also to the critical chemical
potential. Equation ~(\ref{rel5}) was used for
the finite-size scaling analysis. We found that
instead of extrapolating data for each value of $U$
to the $L \to \infty $ limit independently,
a much more efficient procedure is to perform a joint
finite-$L$ and finite-$U$ analysis.
To this end we heuristically introduce parameters accounting
for non-universal finite-$U$ corrections by adding linear in
$U$ terms to each of the three of the dimensionless constants:
$A \to A+BU$, $A' \to A'+B'U$, and $A'' \to A'' + B''U $.
We thus have six fitting parameters to describe
all our data points \cite{fit}. The data for $n_c(U,L)$ and
$\mu_c (U,L)$ are presented in Fig.~1. The fitting procedure
yields $A=(6.07 \pm 0.05)\cdot 10^{3}$, $A_{\mu}=(211 \pm 6)$,
which, according to Eq.~(\ref{xi_A}), means that
\begin{equation}
\xi=380 \pm 3 \; , ~~~~~ \xi_{\mu}=13.2 \pm 0.4 \; .
\label{xi}
\end{equation}
The fit is extremely good---20 points for the critical density at
$U\le 2.5$ and $L m (UT)^{1/2} >15 $, each calculated with
relative accuracy of order $10^{-4}$, are described with the
confidence level of $62$ \%.

\vspace*{-0.3cm}
\begin{figure}
\epsfxsize=0.48\textwidth
\hspace*{-0.5cm} \epsfbox{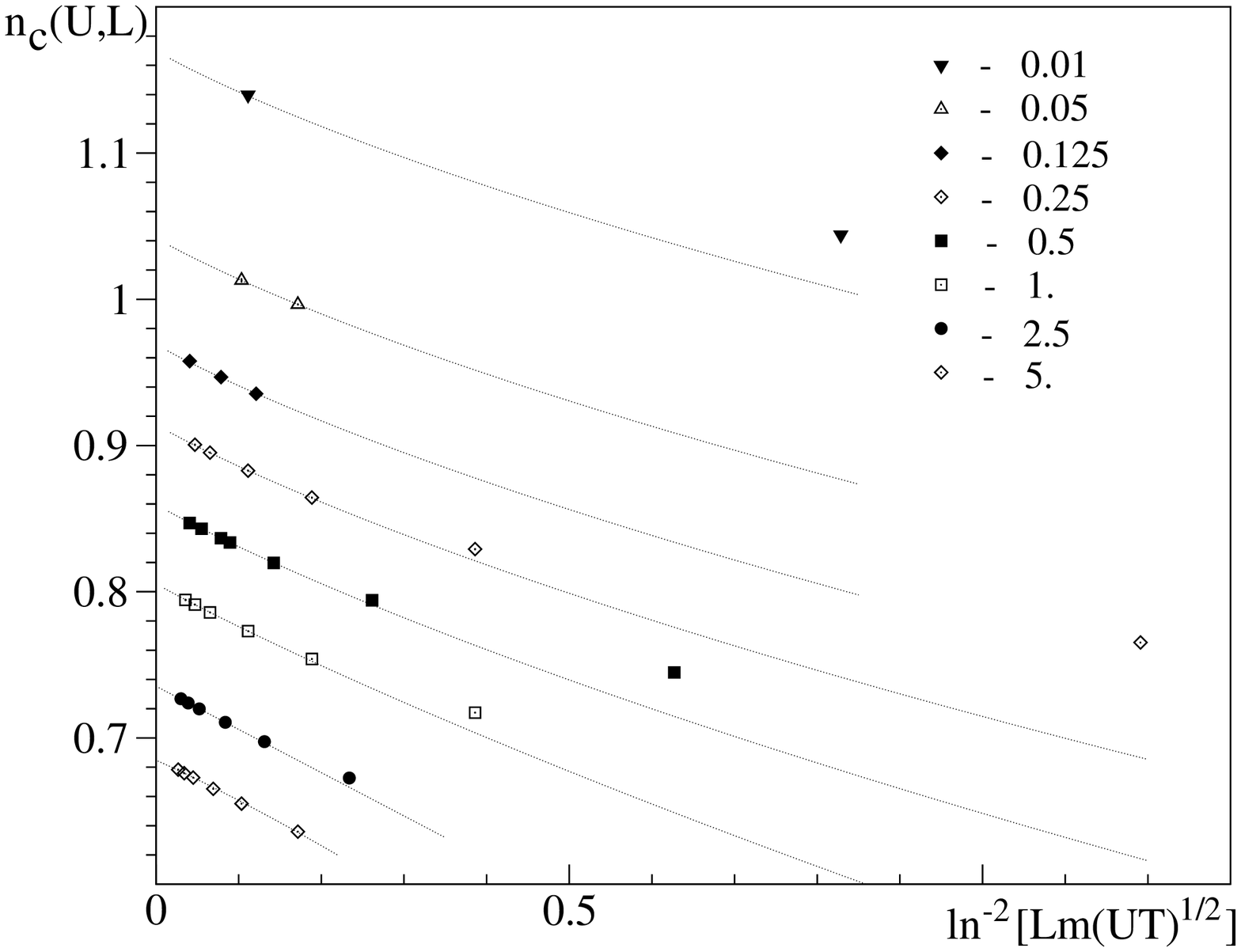}
 \end{figure}
\vspace*{-5.5cm}
\begin{figure}
\epsfxsize=0.48\textwidth
\hspace*{-0.5cm} \epsfbox{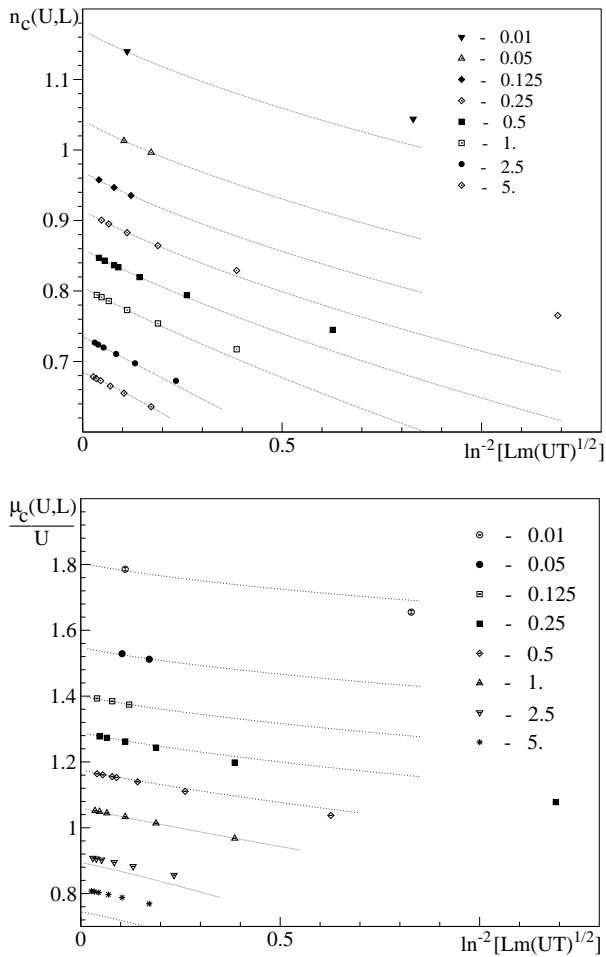}
\vspace*{-3.8cm}
\caption{Critical density and chemical potential
for various coupling parameters
and system sizes.
Typical error bars are  much smaller
than symbol sizes. The dotted line is the fitting function
described in the text.}
\label{fig:fig1}
\end{figure}

Experiments on helium films often report that the ratio
$n_s(T_c)/n_s(0) = n_s(T_c)/n_c = 2mT_c/\pi n_c $ is close to
$0.75$ \cite{Reppy,Hallock}. Our simulation predicts that this
ratio is given by
\begin{equation}
n_s(T_c)/n_c = {4 \over 5.94 - \ln (mU) } \; ,
\label{ratio1}
\end{equation}
and $mU \approx 1.8 $ is required to describe helium films,
provided the small-$U$ approximation may be pushed that far
\cite{comment}. We are not aware of the published data on the
critical chemical potential. [For helium and hydrogen films on
substrates one has to shift $\mu_c$ by the value of the absorption
energy (for the delocalized atom, in the case of helium film),
$\mu_c \to \mu_c=E_0 + (m T U/\pi ) \ln ( \xi_{\mu}/ mU )$. In
thermal equilibrium this quantity can be readily measured through
the chemical potential of the bulk vapor.]

In the absence of long-range order parameter, 2D systems below
$T_c$ are characterized by the local correlation properties of the
quasicondensate density, identical to those of a system with
genuine condensate \cite{KSS}. These properties reflect the
specific structure of the $\psi$-field:
\begin{eqnarray}
\psi ({\bf r})  & = & \psi_0 ({\bf r}) + \psi_1 ({\bf r}) \; ,
\label{psi} \\
\psi_0 ({\bf r})& \approx &  \sqrt{n_0} \, e^{i \Phi ({\bf r})} \; ,
\label{psi_0}
\end{eqnarray}
where the quasicondensate density $n_0$ may be considered as a constant,
and $\psi_1$ is the Gaussian field independent of $\psi_0$. Both experiment
\cite{Safonov} and model Monte Carlo simulations \cite{KKKPS}
indicate that in 2D systems with $mU \sim 1$ the
quasicondensate correlations appear well above $T_c$
and are pronounced at $T_c$. Below we show that this is a generic feature
of weakly interacting $|\psi|^4$-models.

It is convenient to characterize the quasi-condensate properties
by the correlator
\begin{equation}
Q =  2\langle \, |\psi|^2\rangle^2  - \langle \, |\psi|^4 \rangle
\; . \label{Q}
\end{equation}
The Gaussian component of the field obeys the Wick's theorem and
does not contribute to Eq.~(\ref{Q}). If, for a moment, by
$\psi_1$ we understand short-wave harmonics of $\psi$, we conclude
that only long-wave and strongly non-linear harmonics with the
momenta $k \sim k_c$ contribute to the correlator $Q$, i.e. $Q
\sim \tilde{n}^2$. Thus, we expect that all weakly interacting
$|\psi|^4$-models satisfy
\begin{equation}
Q = C_* m^2 T^2 ~~~~~~~(T=T_c)  \label{C_*}
\end{equation}
in the limit of small $U$, where $C_*$ is a universal constant. By
definition, $n_0=\sqrt{Q}$.

The finite-size and small-$U$ analysis of the data for $Q(U,L)$
was done in complete analogy with previously discussed cases of
$n_c(U,L)$ and $\mu_c(U,L)$ (see Ref.~\onlinecite{fit}). We found
that
\begin{equation}
C_* = 1.30 \pm 0.02
 \label{C_*_result}
\end{equation}
The ratio between $n_0(T=T_c)$ and $n_c$ describes how pronounced
are the quasicondensate correlations in the Bose gas at the BKT
point:
\begin{equation}
{n_0^{(T=T_c)} \over n_c} =  {2\pi \sqrt{C_*} \over \ln (\xi /
mU)} = {7.16 \over 5.94 + \ln (1/mU)}  \; . \label{ratio}
\end{equation}
We see, that it is of order unity unless $mU$ is exponentially small.
Another interesting ratio is
\begin{equation}
{n_0 \over n_s} =  {\pi \sqrt{C_*} \over 2} \approx 1.79 ~~~~~~~
(T=T_c) \;,
 \label{ratio_2}
\end{equation}
which is interaction independent and shows that the superfluid density is
substantially smaller than the quasicondensate density at $T_c$.

\vspace*{-0.4cm}
\begin{figure}
\epsfxsize=0.48\textwidth \hspace*{-0.5cm} \epsfbox{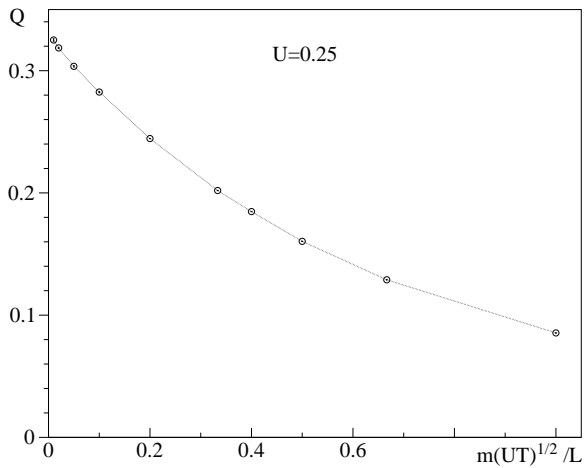}
\vspace*{-3.8cm} \caption{Quasicondensate correlations as a
function of system size. The dotted line is to guide the eye.}
\label{fig:fig3}
\end{figure}

Finally, we would like to derive an accurate estimate  for the
mode-coupling radius $r_c$. In an ideal system $Q \equiv 0$.
Hence, $Q(L)$ should decrease with decreasing $L$, and for system
sizes $L \sim r_c$ it has to drop significantly from its
thermodynamic value. We rather formally define $r_c$ from
$Q(L=r_c) \approx Q(L \to \infty) / 2$, and from Fig.~2 obtain
\begin{equation}
r_c \approx 2 / m (UT)^{1/2} \; .
 \label{r_c}
\end{equation}

We conclude by noting that Nelson-Kosterlitz formula (\ref{NK})
and Eqs.~(\ref{n_c_quantum}), (\ref{mu_c}), and (\ref{C_*})
constitute a complete set of equations which allow to fully
determine system parameters from measurements with independent
cross-checks. We are not aware of another study were dimensionless
constants $\xi$, $\xi_{\mu }$, and $C_*$ were determined with high
precision.

We thank J. Machta and R. Hallock for valuable discussions.
This work was supported by the National Science Foundation under Grant
DMR-0071767. BVS acknowledges a support from Russian Foundation for
Basic Research under Grant 01-02-16508.

\end{multicols}


\begin{thebibliography}{99}
\bibitem{fit} Fitting parameters are obtained from the
stochastic optimization procedure. The error bars are estimated
from fluctuations observed when some of the points are
added/removed from the optimization. We have also tried to look
for the non-universal finite-$U$ corrections of the form $U \ln
U$, but, within the error bars, obtained the same result for
$\xi$, and $\xi_{\mu}$. On another hand, $U \ln U$ corrections
were important in the fit for $Q(U,L)$.
\bibitem{comment} Actually, the interaction $mU \sim 2$ is too
large for a quantum Bose system to be accurately described as a
weakly interacting gas. From our estimate for the mode-coupling
radius, Eq.~(\ref{r_c}), we see that at such interactions $r_c$ is
already on the order of the interparticle distance. Also,
Eq.~(\ref{ratio}) makes no sense at all for $mU >1$.
\end{thebibliography}
\end{document}